\documentstyle[aps,prb,cite,multicol]{revtex}

\input{psfig}

\begin{document}

\title{Optical absorption in semiconductor quantum dots:\\ Nonlocal effects}

\author{F. Thiele, Ch. Fuchs, and R. v. Baltz}

\address{Institut f\"ur Theorie der Kondensierten Materie, 
Universit\"at Karlsruhe,\\%
D-76128 Karlsruhe, Germany}

\date{\today}

\maketitle

\begin{abstract}
The optical absorption of a single spherical semiconductor quantum dot 
in an electrical field is studied taking into account the nonlocal 
coupling between the electric field of the light and the polarizability 
of the semiconductor.
These  nonlocal effects lead to a small size and field dependent shift and 
broadening of the excitonic resonance, which may be of interest in future
precision experiments. 

\end{abstract}

\begin{multicols}{2}

\section{INTRODUCTION}

In recent years, optical investigations on semiconductor
quantum dots (QDs) have attracted much attention both from the physical
and technological points of view.
Such quasi zero dimensional systems display a discrete atomic--like
density of states which is essentially different from that of higher 
dimensional systems,  see e.g. textbooks\cite{Ivch-Pikus,Haug} 
and reviews\cite{Banyai,Woggon,Gaponenko}.
In order to realize the full potential of QDs, it is very important to 
understand the basic physical properties of single QDs whose fabrication
and spatially resolved optical spectroscopy became possible in the last decade,
see e.g.\ \cite{Brunner,Asaoka,Kulak}. 

Optical properties of semiconductor microstructures are usually described
in the approximation of a homogeneous effective medium, which is applicable
as long as the optical wave length in the structure is much larger than
the relevant length scales in the system. This is a well accepted and frequently
stated condition, e.g.\ Eq.(4) of Ref.\cite{Ivchenko}. 
Near a sharp excitonic resonance, however, this condition is violated and
the homogeneous medium approximation breaks down. 
Hence, the nonlocal character
of the relation between the electrical field of the light and the polarization
of the QD comes into play. In particular, this holds if the dimensions 
of the QD become comparable with the exciton Bohr radius, see e.g.\ 
the discussion by Schmitt--Rink et al.\  \cite{Schmitt}. 

In the bulk of a semiconductor, radiative decay of an exciton is forbidden
by conservation of momentum.
In a confined structure, however, such  decay processes become possible and 
they are considered to be relevant to estimate the long intrinsic radiative
lifetimes of excitons \cite{Citrin}.
The nonlocal coupling leads to an additional size and polarization dependent 
line--shift and line width--which contributes to the radiative decay 
of an exciton in a confined system and, thus, should be taken into account
in the analysis of precision experiments. Ultranarrow  ($30 \dots 60\mu eV$) 
photoluminescence lines have been reported\cite{Brunner,Asaoka} which, 
presumably, are still broader than the intrinsic line--width.
We expect that the width of absorption lines is of the same order. 
However, with the exception of Refs.\cite{Ivchenko,Ivch-Kav}, 
the nonlocal coupling between the polarization and the electric field 
of the light wave is generally omitted without giving reasons.

The physical motive of our paper is to
formulate a sound theoretical basis of optical absorption (extinction)
in a confined system with the inclusion of nonlocality and to give a 
reliable prediction of the  corrections to the line--shift and line width.
Following Ivchenko and Kavokin\cite{Ivch-Kav}, we solve 
the scattering problem of an incident electromagnetic wave on
a QD with inclusion of the spatial dependence
of the polarizability in the vicinity of a single excitonic resonance line.
In addition, we consider the influence of a (symmetry breaking) 
static electric field which uncovers nonlocal effects, Sec. II. 
Sec. III contains our numerical results  and discussion
for a $CdSe$ QD in an external electric field.
Unexpectedly, the numerical values are rather small 
but the field induced line shift is of the same order of magnitude 
as the experimental line--width without field.
Hence, the reported corrections may be relevant in future  precision experiments.

\section{THEORY OF NONLOCAL OPTICAL ABSORPTION}
\subsection{Nonlocal Susceptibility}

In linear response theory the relation between the polarization and electric
field reads\cite{Keldysh} 
\begin{equation}
{\mathbf P}({\mathbf r},\omega)=
    \int{ \bbox{\chi}  ({\mathbf r},{\mathbf r}';\omega)\, 
          {\mathbf E}({\mathbf r}',\omega) \,d^3r'} \, .
\label{nl-sus}
\end{equation}
In a local approximation 
$\bbox{\chi} ({\mathbf r},{\mathbf r}' ; \omega)=
\delta({\mathbf r} - {\mathbf r}') \bbox{\chi} (\omega)$.
Nonlocality has different origins in the bulk and in a QD.
In the homogeneous bulk $\bbox{\chi}$ depends only on the relative
coordinate ${\mathbf r}_1 = {\mathbf r} - {\mathbf r}'$
so that Eq.(\ref{nl-sus}) becomes in Fourier--space
${\mathbf P} ({\mathbf q},\omega)= \bbox{\chi} ({\mathbf q},\omega) 
{\mathbf E} ({\mathbf q},\omega)$.
The wave number dependence of $\bbox{\chi} ({\mathbf q},\omega)$
is usually termed spatial dispersion and it originates from the delocalized
nature of the excitations. 
On the other hand, in a QD $\bbox{\chi}$ depends on both
${\mathbf r}$ and ${\mathbf r}'$ and nonlocality is governed by the confinement
of the wave functions, see Eq.\ (\ref{chi-nonloc}).

Near resonance the contribution of a single exciton line to the
polarizability  is given by \cite{Haug,Keldysh} 
 \begin{eqnarray}
  \nonumber  
  \chi_{\alpha\beta}({\mathbf r},{\mathbf r}';\omega)= 
  \left(\frac{e}{m_0}\right)^2
  \frac{1}{\hbar\omega_0^2}
  \frac{ p_\alpha^{cv} p_\beta^{cv} } {\omega_0-\omega-i\Gamma}\\
  \times\Phi({\mathbf r},{\mathbf r})\, \Phi^*({\mathbf r'},{\mathbf r'}) \, ,
  \label{chi-nonloc}
\end{eqnarray}
where, $\alpha,\beta\in\{x,y,z\}$, $\omega_0$ is the exciton transition 
frequency, $\Gamma$ is a phenomenological damping rate, 
${\mathbf p}_{cv}$ is the
interband momentum matrix element, and $\Phi({\mathbf r}_e,{\mathbf r}_h)$ 
denotes the exciton envelope--function.
In addition, we neglect optical anisotropy and approximate 
$\chi_{\alpha,\beta}$ by a  diagonal matrix. As a result, we have
\begin{eqnarray}
\label{ex-pol}
{\mathbf P}_{{\mathrm ex}} ({\mathbf r},\omega)&=&
 \epsilon_0 T(\omega)\,\Phi({\mathbf r})\bbox{\Lambda}(\omega)\, ,\\
\label{Lambda}
  \bbox{\Lambda}(\omega) &=&
 \int{ d^3r' \, \Phi^*({\mathbf r}') \,{\mathbf E}({\mathbf r}',\omega)}\, ,\\
  T(\omega)&=& T_0 \frac{\omega_0}{\omega_0-\omega-i\Gamma}\, , \\
  T_0 &=&\frac{e^2 |p_{cv} |^2}{\epsilon_0 m_0^2\,\hbar\,\omega^3_0}\, .
\label{T}
\end{eqnarray}
For shortness, $\Phi({\mathbf r}):=\Phi({\mathbf r},{\mathbf r})$.
Clearly, the length scale of nonlocality is set by the exciton envelope 
function, and nonlocal corrections are expected 
to be particularily important for small QDs.

\subsection{Scattering Problem}

We consider a linearily polarized monochromatic electromagnetic wave
propagating along the $x$--axis which is scattered by the QD centered at the
origin. Following Jackson\cite{Jackson}, the electric field is represented by 
\begin{eqnarray}
\label{Etot}
  {\mathbf E} ({\mathbf r},\omega) &=& 
  {\mathbf {\hat e}}_{\eta} E_i e^{ikx}+ {\mathbf E}_{s} 
   ({\mathbf r},\omega)\, , \\
\label{Escatt}
  {\mathbf E}_s ({\mathbf r},\omega) &=& \frac{ k_0^2}{\epsilon_0} 
   \int \!\! d^3\!r' \, {\mathbf G} ({\mathbf r} - {\mathbf r}')\,
                        {\mathbf P}_{{\mathrm ex}} ({\mathbf r}',\omega)\, .  
\end{eqnarray}
${\mathbf{\hat e}}_{\eta}$ is the unit polarization vector in 
direction $\boldmath \eta$, $k_0=\omega/c$, $k^2=\epsilon_b k_0^2$,
$\epsilon_b$ is a ``background'' dielectric constant, which accounts for all 
non--resonant contributions at higher frequencies to the polarization 
which are not contained in Eq.(\ref{chi-nonloc}), and
$\mathbf G$ is the matrix Green function of the wave--equation
\begin{eqnarray}
 \label{Green}
  G_{\alpha\beta} ({\mathbf r})=\left(\delta_{\alpha\beta}+\frac{1}{k^2}
                \frac{\partial^2}{\partial r_{\alpha} \partial r_{\beta}}\right)
                \frac{e^{ikr}}{4\pi r}  \nonumber\\
             =\frac{e^{ikr}}{4\pi r}\left[ \frac{2}{3}\delta_{\alpha\beta}+
          \left( 3\frac{r_{\alpha}r_{\beta}}{r^2}-\delta_{\alpha\beta}\right)
          \right.          
          \nonumber\\
          \left.
          \times\Bigl(\frac{1}{(kr)^2}-\frac{i}{kr}-\frac{1}{3}\Bigr)\right]\, .
\end{eqnarray}
$r=|{\mathbf r}|$. ${\mathbf P}_{{\mathrm ex}}$ 
implicitly depends on the unknown ${\mathbf E}$ 
field, Eq. (\ref{ex-pol},\ref{Lambda}), but, fortunately, the vector 
$\bbox{\Lambda}(\omega)$, can be obtained 
directly\cite{Ivch-Kav}. Multiplying Eq. (\ref{Etot}) by
$\Phi^*({\mathbf r})$ and integrating over ${\mathbf r}$, 
we obtain a linear vector equation
\begin{equation}
\label{Lambda-eq}
   \bbox{\Lambda} (\omega) = 
   {\mathbf{\hat e}}_{\eta} E_i \Phi^*(k{\mathbf{\hat e}}_x) + 
   \bbox{\Xi} (\omega)\, \bbox{\Lambda} (\omega)
\end{equation}
which can be solved by matrix inversion
\begin{equation}
\label{Lambda-sol}
   \bbox{\Lambda} (\omega)=E_i\,\Phi^*(k{\mathbf{\hat e}}_x)
   \left[ {\mathbf I}- \bbox{\Xi}(\omega) \right]^{-1}\, 
   {\mathbf{\hat e}}_{\eta}\, .
\end{equation}
$\Phi({\mathbf k})$ is the Fourier--transform of  
$\Phi({\mathbf r})$, $\bbox{\Xi}$  is a $3\times 3$ matrix,
\begin{eqnarray}
   \label{phi-FT}
   \Phi({\mathbf k}) &=&\int { \! d^3\!r \, \Phi({\mathbf r})\,  
    e^{-\imath{\mathbf k}{\mathbf r}} }\, ,\\
   \label{xi-def}
    \bbox{\Xi} (\omega) &=& k_0^2 \, T(\omega)\, {\mathbf K}\, ,\\
    {\mathbf K} &=& 
   \int{\! d^3\!r \int \! d^3\!r'\, 
    {\mathbf G} ({\mathbf r}-{\mathbf r}') \,
     \Phi^*({\mathbf r})\, \Phi({\mathbf r}') }\, ,
\label{K-def}
\end{eqnarray}
and ${\mathbf I}$ is the $3\times 3$ unit matrix.
Eq.\ (\ref{Lambda-sol}), together with 
Eqs.\ (\ref{ex-pol},\ref{Etot},\ref{Escatt}) 
completes the solution of the electromagnetic scattering problem. 
The magnetic field of the wave is given by 
${\mathbf H}=-\imath\nabla\times{\mathbf E}/(\mu_0 \omega)$. 
Large nonlocal corrections are expected to occur if 
${\mathrm det}\left[ {\mathbf I}- \bbox{\Xi}(\omega) \right]\approx 0$.
(This condition is equivalent to Eq.(4) in Ref.\cite{Ivchenko}).

\subsection{Optical Absorption}

The optical absorption is determined by the time--averaged
energy flux ${\bar{\mathbf S}}$
through a closed surface centered around the QD, where
${\mathbf S}={\mathbf E}\times{\mathbf H}$. 
${\bar{\mathbf S}}={\bar{\mathbf S}}_i+{\bar{\mathbf S}}_s+
{\bar{\mathbf S}}_{{\mathrm ext}}$
can be decomposed in an incident, scattered, and extinction 
contribution\cite{Jackson,Bohren}.
In addition, we assume that the QD is surrounded by a nonabsorbing medium 
of the same background dielectric constant. 
As a result, the net absorbed energy flux becomes
\begin{equation}
  W_a =-\oint_A {\bar{\mathbf S} } \!\cdot\! {\mathbf{\hat e}}_r \,dA = 
             W_i - W_s + W_{{\mathrm ext}}\, .
\end{equation}
$W_i=0$, whereas $W_s, W_{{\mathrm ext}}>0$.
$W_{{\mathrm ext}}=W_a + W_s$ gives the missing energy flux out 
of the incident wave.

To calculate the optical absorption  of the QD
we only need the scattered field in the far field, $r\gg R\ge r'$.
In leading order, we obtain
\begin{eqnarray}
  \label{E-scatt}
  {\mathbf E}_s ({\mathbf r},\omega)&=&E_i\, \frac{e^{ikr}}{4\pi r}\,
  {\mathbf F}({\mathbf{\hat e}}_r,\omega)\, ,\\  
  \label{s-amp}
  {\mathbf F}({\mathbf{\hat e}}_r,\omega)&=&- k_0^2 \, T(\omega)\,
   \Phi^*(k{\mathbf{\hat e}}_x)\,\Phi(k{\mathbf{\hat e}}_r)\,
   {\mathbf{\hat e}}_r\times \left[ {\mathbf{\hat e}}_r \times 
   \bbox{\xi} \right]\, ,\\
   {\mathbf H}_s ({\mathbf r},\omega)&=&
   \sqrt{\frac{\epsilon_0\epsilon_b}{\mu_0}}\,
   {\mathbf{\hat e}}_r\!\times\!
  {\mathbf E}_s ({\mathbf r},\omega)\, ,\\
   \bbox{\xi}(\omega) &=& 
   \left[{\mathbf I}-\bbox{\Xi}(\omega) \right]^{-1} 
  {\mathbf{\hat e}}_{\eta} \label{xi}\, .
\end{eqnarray}
${\mathbf F} ({\mathbf{\hat e}}_r,\omega)$ 
is the scattering amplitude of the wave in direction 
${\mathbf r}$. Eqs.(\ref{E-scatt}-\ref{xi}) can be considered
as a nonlocal generalization of Mie--scattering. 

The scattered energy flux
is easy to obtain as the respective electric and magnetic fields are transversal 
\begin{eqnarray}
    \label{Wscatt}
  W_s = I_i \left | \frac{k_0^2\, T(\omega)}{4\pi}\, 
     \Phi(k{\mathbf{\hat e}}_x)\right |^2 \nonumber\\
     \times 
     \int \left| \Phi(k{\mathbf{\hat e}}_r)\right|^2 \left( |\bbox{\xi}|^2
    -|{\mathbf{\hat e}}_r\cdot \bbox{\xi}|^2 \right) \, d\Omega_r \,.
\end{eqnarray}
$I_i=\sqrt{\epsilon_0\epsilon_b/\mu_0}\, E_i^2/2$ 
is the intensity of the 
incident light wave and $d\Omega_r$ is the surface element of the unit sphere. 
For  $kR\ll 1$ the integration is trivial so that 
Eq.\ (\ref{Wscatt}) simplifies to 
\begin{equation}
  \label{eq:Ws_erg_app}
   W_s=I_i\, \frac{k_0^4}{6\pi} \left| T(\omega)\, 
            \bbox{\xi}(\omega)\, \Phi(k{\mathbf{\hat e}}_x)\,
             \Phi({\mathbf k}=0)\right|^2 \, .
\end{equation}

The extinction is conveniently calculated from the optical 
theorem which relates $W_{{\mathrm ext}}$
to the imaginary part of the scattering amplitude
in forward direction\cite{Jackson, Bohren}
\begin{eqnarray}
   W_{{\mathrm ext}} &=& \frac{I_i}{k} \Im{\, {\mathbf{\hat e}}_{\eta}\, 
             {\mathbf F}({\mathbf{\hat e}}_x,\omega)}
             \nonumber\\
             &=&
             \frac{I_i}{k} |\Phi(k{\mathbf{\hat e}}_x)|^2\,
             \Im \left\{k_0^2\, T(\omega)\, 
             \left( \left[{\mathbf I}-\bbox{\Xi}(\omega)
                        \right]^{-1} \right)_{\eta\eta}\right\}\, . 
  \label{Wext}
\end{eqnarray}

For typical material parameters  
($E_g\approx$ 2 eV, $R\approx 5$nm, $\Gamma=0.1\dots1$meV)
$|\Xi_{\alpha\beta}(\omega_0)|\lesssim 0.1$ so that an expansion of
$({\mathbf I}-\bbox{\Xi})^{-1}$ is reasonable. 
For a rough estimate, we consider the zeroth--order and assume $kR\ll 1$,
where $\bbox{\xi}={\mathbf{\hat e}}_{\eta}$ so that 
$W_s/W_{{\mathrm ext}}\approx 10^{-4}$,
i.e. the optical absorption is dominated by the extinction. 
Our numerical results  indicate that this 
estimate holds for other reasonable parameters, too.

Up to first order 
$[{\mathbf I}-\bbox{\Xi}]^{-1}=
{\mathbf I}+\bbox{\Xi}+O(\bbox{\Xi}^2)$ 
the optical cross section 
$\sigma_a=\sigma_{{\mathrm ext}}=W_{{\mathrm ext}}/I_i$ 
is given by
\begin{eqnarray}
 \nonumber
  \sigma_a (\omega,R) = \frac{\omega_0}{c\sqrt{\epsilon_b}}\,T_0 \, 
  \left|\Phi({\mathbf k}=0)\right|^2\,\\
  \times
  \Im \left\{\frac{\omega_0}{(\omega_0+\Delta\omega_0)-\omega-
   \imath (\Gamma+\Delta\Gamma)}\right\}\,,
 \label{cross-sec}\\
 \Delta\omega_0 -\imath\Delta\Gamma = -k_0^2 T_0 K_{\eta\eta}\,.
\label{do-dg}
\end{eqnarray}
$\Delta\omega_0$ and  $\Delta\Gamma$ respectively denote the shift and 
broadening of the excitonic resonance which are caused by the nonlocality, 
and $\left|\Phi({\mathbf k}=0)\right|^2$ 
is a measure of the oscillator strength of the excitonic line.

For the first order nonlocal correction of the line--width and line--shift, 
only the diagonal elements of Eq.\ (\ref{K-def}) are needed. 
According to Eq.\ (\ref{Green}) (second line), the Green function 
is made up of an isotropic and a traceless term,
${\mathbf G}={\mathbf G}^{(1)}+{\mathbf G}^{(2)}$. 
Therefore, $K_{\eta\eta}=K^{(1)}+K^{(2)}_\eta$, where $K^{(1)}$ is independent 
of the polarization $\eta$ and $K_x^{(2)}+K_y^{(2)}+K_z^{(2)}=0$.
Moreover, in the ground state we still have rotational symmetry 
around the $z$--axis,  thus $K_x^{(2)}=K_y^{(2)}=-K_z^{(2)}/2$, and, 
in addition, $K_{\alpha\beta}=0$, $\alpha\not=\beta$, so that the inversion of 
$[{\boldmath I} - \bbox{\Xi}]$, Eq.(\ref{Wext}), is trivial.
For $F_z=0$ the ground state exciton wave function is 
isotropic which implies  $K_\eta^{(2)}=0$ for all $\eta$.

\subsection{Exciton States}

In a spherical QD with infinite confinement the electron/hole Hamiltonians 
(without external field) read
\begin{equation}
  \label{e/h-ham}
  H_j=-\frac{\hbar^2}{2m^*_j} \Delta \,,
\end{equation}
where $r\leq R$, j=e/h, and $m^*_j$ are the respective effective masses.  
The eigenstates of $H_j$ (omitting the indices j=e/h)
are well known from textbooks, e.g. Ref.\cite{Cohen}
\begin{eqnarray}
  \label{e/h-wavef}
  \psi^{(0)}_{nlm}({\mathbf r})&=&F_{nl}(r)\, Y^m_l(\theta,\varphi)\, ,\\
         F_{nl}(r) &=&\sqrt{\frac{2}{R^3}}\,\frac{j_l(a_{ln}\frac{r}{R})}
             {j_{l+1}(a_{ln})}\, ,\\
  \label{e/h-energies}
          E^{(0)}_{nl} &=& \frac{\hbar^2 a_{ln}^2}{2m^* R^2}\, . 
\end{eqnarray}
Electron/hole energies count from their respective band edges,
quantum numbers $n,l,m$ have their usual meaning and
$a_{ln}$, $n=1,2,\dots$ denotes the $n^{th}$ positive zero of the 
spherical Bessel function $j_l(x)$\cite{Abramo}.

Next, we consider noninteracting electron--hole pair states which are 
the eigenstates of $H_e + H_h$,
\begin{eqnarray}
 \label{e/h-pair}
  \Psi^{(0)}_{\lambda_e\lambda_h}({\mathbf r}_e,{\mathbf r}_h)
  &=&\psi^{(0)}_{\lambda_e}({\mathbf r}_e)
   \psi^{(0)}_{\lambda_h}({\mathbf r}_h)\, ,\\
\label{pair-en}
  E^{(0)}_{\lambda_e\lambda_h}&=& E^{(0)}_{\lambda_e} + E^{(0)}_{\lambda_h}\, ,
\end{eqnarray}
where $\lambda_j=(n_j,l_j,m_j)$ label the electron/hole states. 
For shortness, the pair states and energies, 
Eqs.\ (\ref{e/h-pair},\ref{pair-en}), will be denoted by 
$|\lambda>$ and $E_p$, where $\lambda=(\lambda_e,\lambda_h)$. 
For a discussion of pair and 
exciton states in QDs see, e.g.\,  Chapter 3 of Ref.\cite{Woggon}.

\begin{figure}
 \centerline{\psfig{file=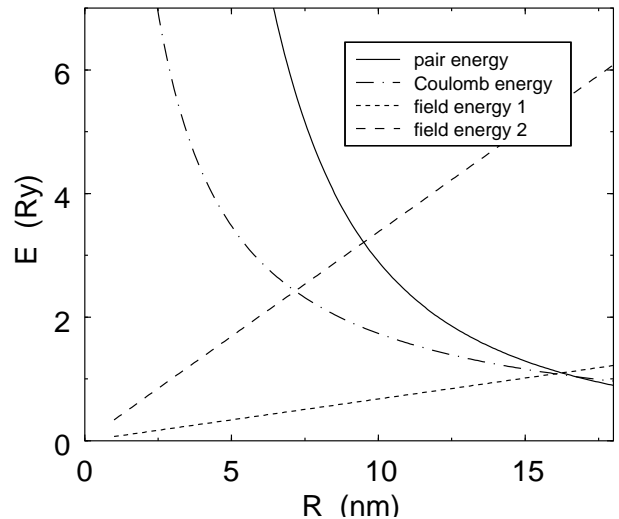,width=8cm}}
 \medskip
\caption{Different energy scales which are set by the confinement: 
pair energy  Eq.(\ref{pair-en}),
Coulomb energy in first--order pertubation theory for $F_z=0$, 
and field--energy, $e F_z R$, for (1) $F_z=1$ and (2) $5$MeV/m.
$Ry^*$ denote the excitonic
Rydberg--energy\cite{Haug}.}
\end{figure}

In the final step we have to include the Coulomb interaction between 
the electron and the hole inside the QD as well as the interaction with
the electrical field (in z--direction) $F_z$.
For QD sizes of $R=2\dots 20$nm the localization energy, Coulomb--energy,
and relevant field--energies $eRF_z$ are approximately of the same magnitude,
see Fig.\  1.

Therefore, we found it convenient to expand the exciton states in the QD 
directly in terms of the pair states Eqs.\ (\ref{e/h-pair}, \ref{pair-en}), 
rather than first constructing exciton states for $F_z=0$.
Total angular momentum of the electrons and holes is not conserved but
there is still rotational symmetry around the $z$--axis so that the z--component
of total electron--hole angular momentum 
$M=m_e + m_h$ is a good quantum number.  
\begin{equation} 
 \label{state-exp}
  |\kappa,M\rangle=\sum_{\lambda} 
  C^{\kappa,M}_{\lambda} |\lambda\rangle\, .
\end{equation}
$\kappa$ is an additional exciton quantum number.
Expansion (\ref{state-exp}) leads to the algebraic eigenvalue problem
\begin{eqnarray}
  \nonumber
  \sum_{\lambda'} C^{\kappa,M}_{\lambda'} \Big( E_{\lambda'}
  \delta_{\lambda\lambda'} +eF_z\langle\lambda|z_e-z_h|\lambda'\rangle \\
 -\frac{e^2}{4\pi\epsilon_0\epsilon_b}
  \langle \lambda|\frac{1}{|{\mathbf r}_e - 
  {\mathbf r}_h |}|\lambda'\rangle  \Big)=
  E^M_\kappa C^{\kappa,M}_{\lambda}\,,
  \label{field-eig}
\end{eqnarray}
where $E^M_\kappa$ is the exciton  energy in a QD 
as measured with respect to the gap.  
(For shortness it will be sometimes referred to by $E_{{\mathrm ex}}$). 
The z-- and  Coulomb matrix elements
can be obtained analytically and some details are listed in 
Appendix  \ref{App-A}. 

\section{Results and Discussion}

Numerical studies have been performed for an optical transition
to the excitonic ground state and the parameters  are appropriate for $CdSe$: 
$m^*_e=0.11m_0, m^*_h=0.44m_0$, $E_g=1.9$eV,
$\epsilon_b\approx 9.8$\cite{Landolt}. 
This implies $T_0= 0.3 \mathrm{nm}^3$ and $Ry= 13$meV. 
For a QD radius of $R=5$nm the pair energy is $E_p=170$meV and the 
exciton binding energy is $E_b= 60$meV so that in total 
the exciton energy (as measured from the gap) becomes 
$E_{{\mathrm ex}}=E_p - E_b = 110$meV. 
Even for the narrowest excitonic lines which have been observed so far, 
$\hbar\Gamma\approx 30 \mu$eV\cite{Brunner,Kulak,Asaoka},
the linear approximation Eq.(\ref{cross-sec}) is still 
reasonable, $|\Xi_{\eta\eta}|\approx 10^{-5} \omega_0/\Gamma < 0.3$.
Although these experiments refer to photoluminescence rather than to
optical absorption similar results for the line--shift and line--width
are expected.

There are two sources which cause the excitonic resonance frequency 
to change with the radius or applied field: 
change of the ``atomic'' transition frequency $\omega_0$  in the dot 
(=difference of energy levels which is not explicitely considered here)
and the radiation induced shift and damping  $\Delta\Gamma$,
$\Delta\omega_0$ which originate from
the (nonlocal) coupling to the electromagnetic field, see Figs.\ 2,3. 
\begin{figure}
 \centerline{\psfig{file=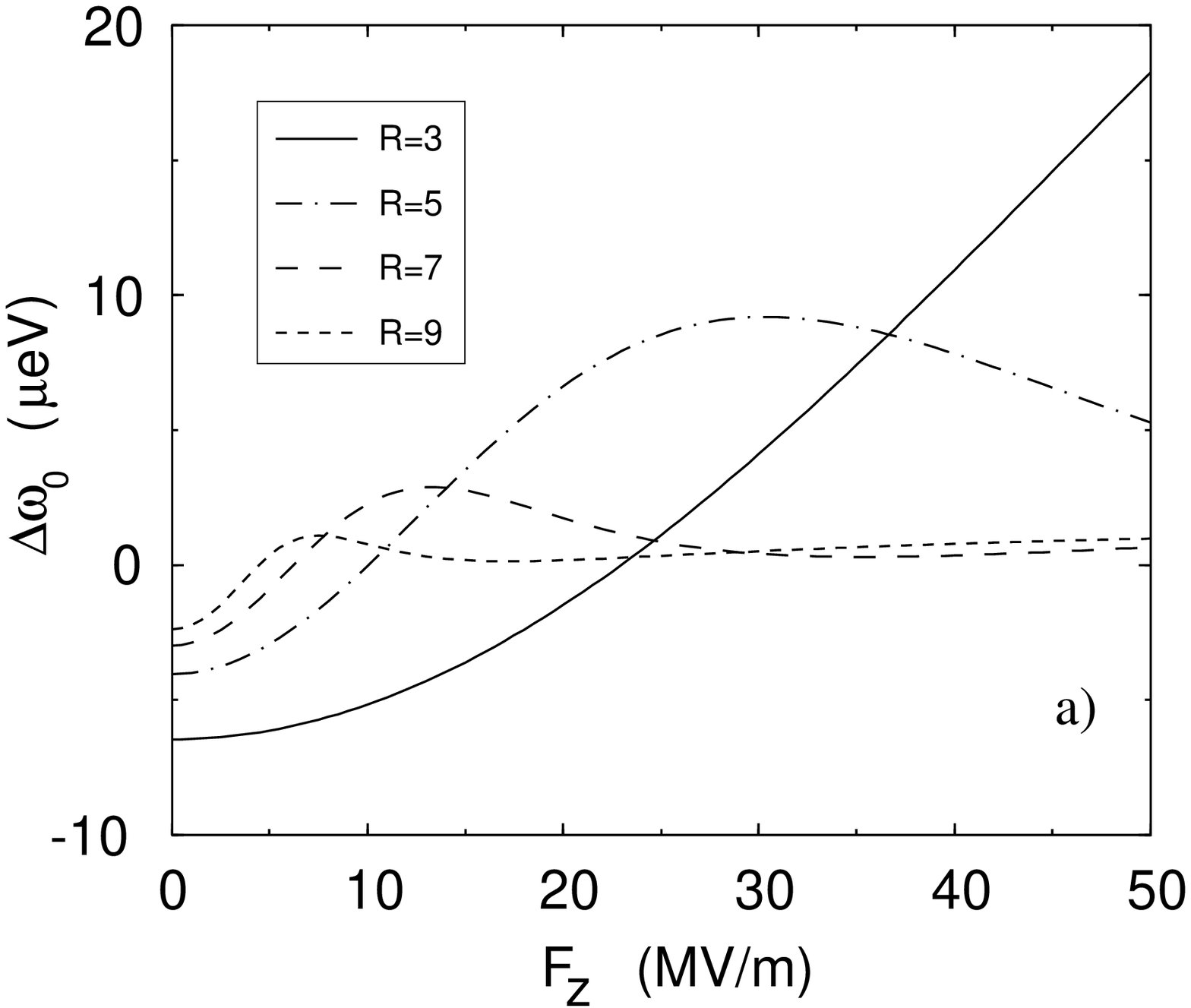,width=7.5cm}}
 \vspace*{0.5cm}
 \centerline{\psfig{file=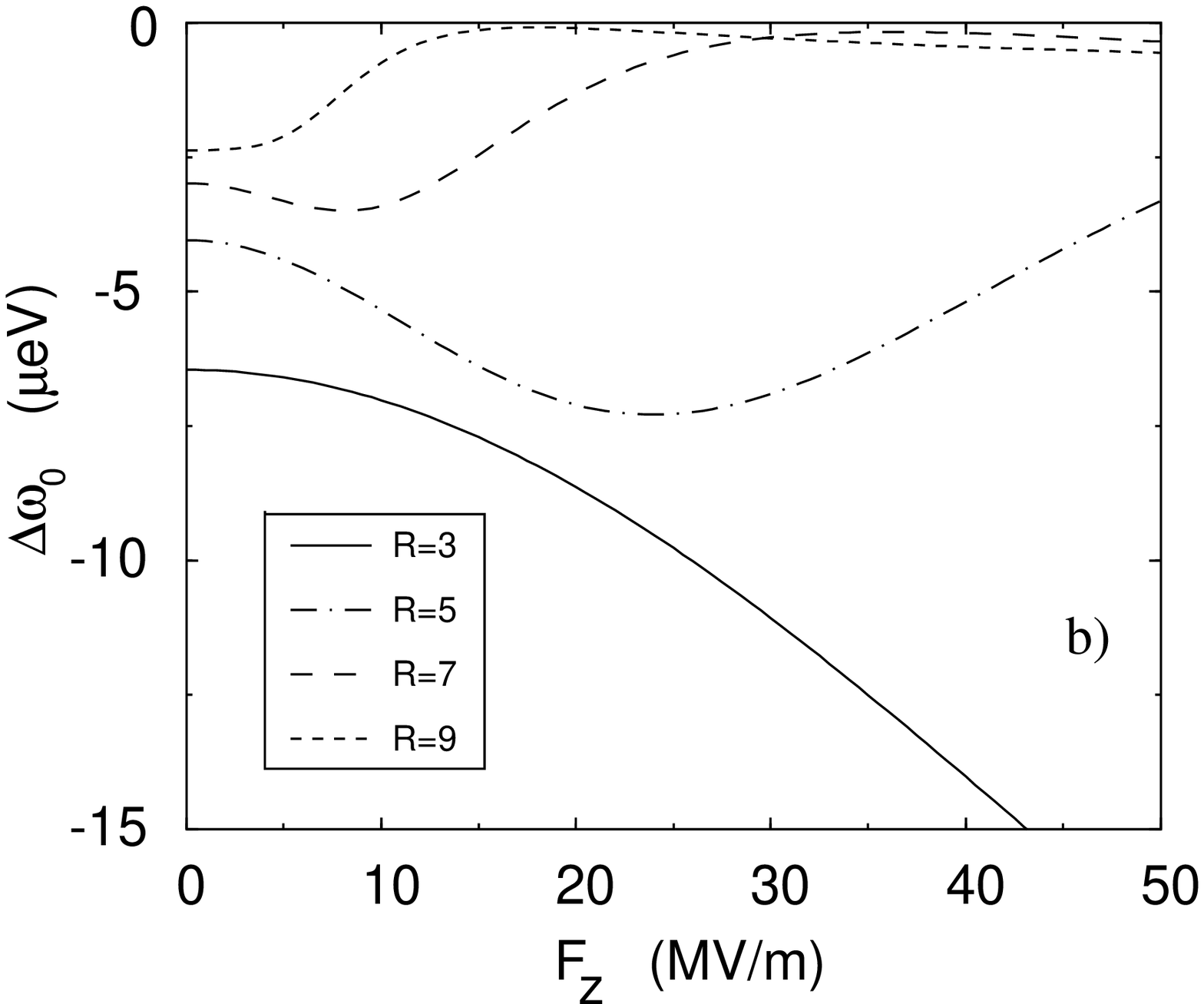,width=7.5cm}}

 \medskip
 \caption{Nonlocal contribution to the exciton line shift.
          (a) Light polarization  parallel  and  (b) perpendicular 
          to the applied field. ($R$ in nm).}
\end{figure}

\begin{figure}
 \centerline{\psfig{file=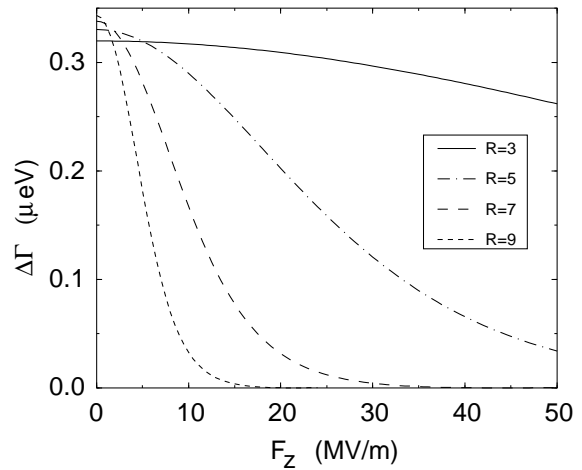,width=7.5cm}}

 \medskip
 \caption{Nonlocal contribution to the exciton line--width as 
          a function of the applied field. 
          (Polarization dependence not resolved. $R$ in nm).}
\end{figure}

\begin{figure}
 \centerline{\psfig{file=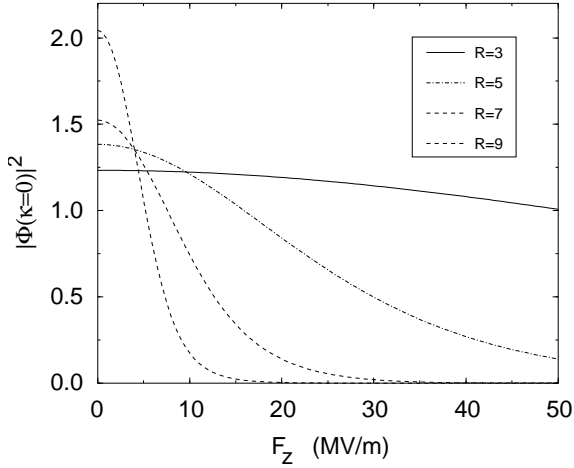,width=7.5cm}}

 \medskip
 \caption{ Relative oscillator strength in the exciton ground state 
           as a function of the applied field ($R$ in nm).} 
\end{figure}
For $F_z=0$: $K_\eta^{(2)}=0$, hence, $\Delta\Gamma$ and $\Delta\omega_0$ 
are both independent of the polarization direction $\eta$.
With applied field,  $K^{(2)}_\eta\not=0$ but
$|\Im K^{(2)}_\eta| \ll |\Im K^{(1)}|$, 
hence $\Delta\Gamma$ is (almost) independent of $\eta$. 
With increasing field, $|\Re K^{(2)}_\eta| > |\Re K^{(1)}|$, so that
$\Delta\omega_0$ displays a pronounced polarization dependence.
For large applied fields, the electron and hole states become spatially 
separated so that both $K^{(1)}$ and $K^{(2)}_\eta$ tend to zero. 
Therefore, $\Delta\omega_0$, $\Delta\Gamma$ as well as the oscillator 
strength, Fig.\ 4, tend to zero.
Therefore, $\Delta\omega_0$ runs over an extremum when
the pair energy approximately equals  field energy, see Fig.\ 1.
Because of the different signs of $K_y^{(2)}$ and $K_z^{(2)}$,
$\Delta\omega_0$ display a maximum/minimum for $y$/$z$ polarization.
   
Eventually we list some general results: 
(i) Smaller dot radii require larger fields 
to generate comparable changes of the electronic states,
(ii) the importance of nonlocal corrections increases with decreasing dot radii
and  deviations from spherical symmetry, 
(iii) the nonlocal contributions are rather insensitive with respect to the 
electron--hole Coulomb interaction, and
(iv) the leakage of the exciton wave function in the host material
may be incorporated using an effective radius $R_{eff}=R+\lambda$,
where $\lambda$ is aproximately the decay length of the (electron)
wave function, $\lambda\approx 1nm$.  

Although the reported influence of the nonlocal coupling between the light
and the polarization of the semiconductor is unexpectedly small, such 
corrections may be of increasing importance for
the analysis of future precision experiments on single QDs,
in particular in an external field or in non--spherical shaped structures.
To the best of our knowledge, the smallest line width reported so far
is that of localized biexciton states in a $GaAs/AlGaAs$ 
structure at low temperatures which is $30\mu$eV\cite{Brunner}. 
The field dependence may come into play for QDs sandwiched between
two metallic leads similar to those which have been fabricated 
recently, e.g.\ \cite{Dekker,Yao}.

\vfill
\section{Acknowledgements}
We thank Prof. E.L. Ivchenko for advice and  many helpful discussions.
Part of this work was supported by the Deutsche Forschungsgemeinschaft through
Sonderforschungsbereich SFB 195.

\begin{appendix}
\section{Matrixelements}
\label{App-A}

The z--matrix elements can be obtained analytically,
\begin{equation}
  \label{z-me}
   \langle n'l'm'|z|nlm\rangle= \langle n'l'|r|nl\rangle 
   \langle l'm|\cos\theta|lm\rangle \delta_{m,m'}\, ,
\end{equation}
where the angular part is listed in $C_X$ of Ref.\ \cite{Cohen}.
\begin{equation}
  \label{r-me}
  \langle n'l'|r|nl\rangle=\frac{2R}{j_{l+1}(\alpha)
    j_{l'+1}(\beta)} I_{l,l'}(\alpha,\beta)
\end{equation}
with abbreviations $\alpha=a_{ln}$, $\beta=a_{l'n'}$, and
\begin{equation}
  \label{Ill'}
  I_{l,l'}(\alpha,\beta)=\int_0^1\!\! d\rho\, 
  \rho^3 \, j_l (\alpha\rho)\, j_{l'}(\beta\rho)\, .
\end{equation}
In particular, for $l'=l+1$, Eq. (\ref{Ill'}) can be rewritten 
by using the recursion formula for $j_{l+1}(x)$ in terms of $j_l(x)$, and 
$j_l'(x)$, as given by (10.1.22) in Ref. \cite{Abramo}
\begin{equation}
I_{l,l+1}(\alpha,\beta)=
        \left( \frac{l}{\beta}+\frac{\partial}{\partial\beta}\right)
        \int_0^1{ \rho^2\, j_l(\alpha\rho)\,j_l(\beta\rho)\,d\rho}\, .
\end{equation}
The remaining integral is a special case of (11.3.29) 
with  (10.1.1) in Ref.\cite{Abramo} when 
$j_l(\alpha)=0$ and $\alpha\not=\beta$ are used.
For $l'=l-1$ we have only to interchange the role of $\alpha, \beta$.

The Coulomb matrix elements are calculated by first expanding 
$|{\mathbf r}_e - {\mathbf r}_h |^{-1}$ in terms of spherical
harmonics\cite{Jackson}, and subsequently performing angular integrations of 
triple products of spherical harmonics in terms of Clebsch--Gordan 
coefficients\cite{Cohen}
\begin{equation}
  \label{Coul-me}
  \bigl\langle\lambda'\bigl|
     \frac{1}{|{\mathbf r}_e-{\mathbf r}_h|} \bigr| \lambda\bigr\rangle= 
     \sum_{l=0}^{\infty} \frac{1}{2l+1} W_{\lambda,\lambda'}\, 
     Z_{\lambda,\lambda'}\, . 
\end{equation}
$W_{\lambda,\lambda'}$ summarizes the result of $r$--integrations, and 
$Z_{\lambda,\lambda'}$ contains a sum on $m$ of products of  
matrix elements of the angular part of the wave--functions.
$Z_{\lambda,\lambda'}$ is only nonzero if 
$l_e+l_e'+l_h+l_h = \mbox{even}$ and 
$\mbox{max} \{|l_e-l_e'|,|l_h-l_h'|\}\leq  l \leq 
 \mbox{min} \{l_e+l_e',l_h+l_h' \}$.
The remaining integrations have been done numerically with 
{\textsc Mathematica}\cite{Wolfram}.

\section{Evaluation of K}
\label{App-B}

Expanding $\Phi({\mathbf r}_e ,{\mathbf r}_h )$ in terms of pair states, 
Eq. (\ref{state-exp}), we obtain
\begin{eqnarray}
  \label{KG}
  K_{\alpha\beta}=\sum_{\lambda_1\lambda'_1\lambda_2\lambda'_2 } 
        C^*_{\lambda_1 \lambda'_1}\, C_{\lambda_2\lambda'_2 }\,
        (\lambda_1\lambda_2|G_{\alpha\beta}|\lambda'_1\lambda'_2 )\, ,\\
        (\lambda_1\lambda_2|G_{\alpha\beta}|\lambda'_1\lambda'_2 )=
  \int\!d^3\!r_1 \int\!d^3\!r_2 \,G_{\alpha\beta}({\mathbf r}_1-{\mathbf r}_2)
  \nonumber\\
  \times
  \psi^*_{\lambda_1}({\mathbf r}_1) \psi_{\lambda_2}({\mathbf r}_2) 
  \psi^*_{\lambda'_1}({\mathbf r}_1)\psi_{\lambda'_2}({\mathbf r}_2)\, .
  \label{def}
\end{eqnarray}
Electron and hole wave functions are  identical, hence the labels
e/h have been dropped. 
The definition used in Eq.(\ref{def}) looks skew but it is useful for
the evaluation in analogy with the Coulomb matrix elements.
The diagonal elements are split into two parts: 
$K_{\eta\eta}=K^{(1)}+K_\eta^{(2)}$, 
where $K^{(1)}$ is independent of the light polarization. 

The numerical evaluation of $K^{(1)}$ is conveniently done by first 
expanding $\exp(ik|{\mathbf r}_1-{\mathbf r}_2|)/|{\mathbf r}_1-{\mathbf r}_2|$
in terms of spherical harmonics\cite{Jackson}, and then follow closely the 
evaluation of the Coulomb matrix elements.
If $kR<0.3$, the expansion of the exponential function in  
$(\lambda_1\lambda_2|G^{(1)}_{\alpha\beta}|\lambda'_1\lambda'_2 )$
leads to fast converging series and, to leading order, we have
\begin{eqnarray}
  \label{G1-me}
  R \left|\Re\left\{ (\lambda_1\lambda_2|G^{(1)}|
    \lambda'_1\lambda'_2 ) \right\}\right| 
    &\lesssim& 10^{-1}\, , \nonumber\\
  R \left|\Im\left\{ (\lambda_1\lambda_2|G^{(1)}|
   \lambda'_1\lambda'_2 ) \right\}\right| 
    &\lesssim& ( \lambda_1\lambda_2|\frac{kR}{6\pi}|
    \lambda'_1\lambda'_2 )\, .
\end{eqnarray}
Therefore, the imaginary part of $K^{(1)}$ is proportional to the 
oscillator strength, Eq(\ref{cross-sec}),
\begin{equation}
  \Im\left\{ K^{(1)}\right\}=\frac{k}{6\pi}
   \left| \Phi({\mathbf k}=0)\right|^2 +  O\!\left((kR)^2\right)\,.
\end{equation}

The evaluation of $K^{(2)}_\eta$ is more difficult than that of $K^{(1)}$ as it
depends on the polarization direction and 
$G^{(2)}_\eta \sim r^{-3}$ is singular at $r=0$. The latter problem,
however,  can be circumvented by doing the angular integrations first.
In addition, the parity of the spherical harmonics, $(-1)^l$, implies that 
the matrix elements of ${\bf G}^{(2)}$ obeys the same angular momentum 
selection rules as ${\bf G}^{(1)}$.
For $kR < 0.3$, we estimate the $G^{(2)}$ matrix elements as
\begin{eqnarray}
R | \Re \left( \lambda_1\lambda_2 | G^{(2)}_\eta |  \lambda_1'\lambda_2'
        \right)  | &\lesssim& \frac{0.2}{(kR)^2}\, ,\\
R | \Im \left( \lambda_1\lambda_2 | G^{(2)}_\eta |  \lambda_1'\lambda_2'
        \right)  | &\lesssim& \frac{0.002}{(kR)^2}\, .
\end{eqnarray}
Therefore,  contributions of  $G^{(2)}$ to $\Im K$ are expected to be small
even in an applied field. For the real part, however, the situation is opposite.
Without the electrical field, the excitonic ground state is mainly 
made up of electron/hole pair states of the type $|n00;n'00>$, with
$\sum |C_{n00;n'00}|^2>0.96$, if $R>5$nm, so that the $G^{(2)}$--elements 
almost vanish  for small fields.
 
The final expression for the numerical evaluation of the $G^{(2)}$--elements
becomes a sum of two five--dimensional integrals, which have been performed 
numerically.
To calculate the optical properties, the lowest $30$  pair states were 
used which lead to  140 independent integrals. 
The estimated numerical accuracy is $0.5$\%.

\end{appendix}

\end{multicols}

\end{document}